\documentclass[12pt,numbers,sort&compress]{article}

\usepackage{natbib}
\usepackage{hypernat}
\usepackage{amsmath}
\usepackage{amsthm}
\usepackage{amsfonts}
\usepackage{amssymb}
\usepackage[all]{xy}
\usepackage{color}
\usepackage[backref,colorlinks=true,linktocpage]{hyperref}

\newlength{\xtrawidth}
\newlength{\xtraheight}
\def\clap#1{\hbox to 0pt{\hss#1\hss}}

{\end{list}}
{\everymath{\displaystyle\everymath{}}\array}%
{\endarray}

\newcommand{\eqdef}{%
  \mathrel{\lower.1mm
    \hbox{$\stackrel{\lower.424ex\hbox{\scriptsize def}}{=}$}}
}
\DeclareMathOperator{\Tr}{Tr}

\DeclareMathOperator{\ad}{ad}

\DeclareMathOperator{\Span}{span}

\newcommand{\dslash}{\mbox{$\partial$ \kern-.92em  \big /}}
\newcommand{\diff}{\mathrm{d}}
\newcommand{\Z}{\mathbb{Z}}

\newcommand{\C}{\mathbb{C}}

\newcommand{\CP}[1]{\mathbb{P}^{#1}}
\newcommand{\IP}[1]{\CP{#1}}

\newcommand{\Rep}[1]{\ensuremath{\mathbf{#1}}}
\newcommand{\barRep}[1]{\ensuremath{\overline{\Rep{#1}}}}

\newcommand{\dual}{\ensuremath{\vee}}

\newcommand{\ZZZ}{{\ensuremath{\Z_3\times\Z_3}}}
\newcommand{\B}[1]{\ensuremath{B_{#1}}}
\newcommand{\dP}[1]{\ensuremath{dP_{#1}}}
\newcommand{\Xt}{{\ensuremath{\widetilde{X}}}}

\newcommand{\Osheaf}{\ensuremath{\mathcal{O}}}

\newcommand{\oB}[1]{\ensuremath{\Osheaf_{\B{#1}}}}
\newcommand{\oXt}{\ensuremath{\Osheaf_{\Xt}}}

\newcommand{\V}[1]{\ensuremath{V_{#1}}}
\newcommand{\Vt}{{\ensuremath{\widetilde{\V{}}}}}
\newcommand{\W}[1]{{\ensuremath{W_{#1}}}}

\newcommand{\Fsheaf}{\ensuremath{\mathcal{F}}}
\DeclareMathOperator{\hd}{hd}

\definecolor{grey}{gray}{0.4}
\newcommand{\Hpq}[3]{\big({#2},{#3}\textcolor{grey}{\big|{#1}}\big)}
\newcommand{\Hst}[4]{\big[{#3},{#4}\textcolor{grey}{\big|{#2},{#1}}\big]}

\newcommand{\cO}{\Osheaf}
\newcommand{\cF}{\Fsheaf}
\newcommand{\cFt}{\widehat{\Fsheaf}}

\newcommand{\atv}{\wedge^2 \Vt}
\newcommand{\tv}{\Vt}
\newcommand{\tx}{\Xt}
\def\z3z3{\ZZZ}

\begin{document}

\begin{titlepage}
  \begin{flushright}
    hep-th/0601204
  \end{flushright}
  \vspace*{\stretch{1}}
  \begin{center}
     \Huge 
      Yukawa Couplings in 
      \\
      Heterotic Standard Models
  \end{center}
  \vspace*{\stretch{2}}
  \begin{center}
    \begin{minipage}{\textwidth}
      \begin{center}
        \large         
        Volker Braun$^{1,2}$, 
        Yang-Hui He$^{3,4}$ and
        Burt A.~Ovrut$^{1}$
      \end{center}
    \end{minipage}
  \end{center}
  \vspace*{1mm}
  \begin{center}
    \begin{minipage}{\textwidth}
      \begin{center}
        ${}^1$ Department of Physics,
        ${}^2$ Department of Mathematics
        \\
        University of Pennsylvania,        
        Philadelphia, PA 19104--6395, USA
      \end{center}
      \begin{center}
        ${}^3$ Merton College, Oxford University,      
        Oxford OX1 4JD, U.K.
      \end{center}
      \begin{center}
        ${}^4$ Mathematical Institute, Oxford, 
        24-29 St.\ Giles', OX1 3LB, U.K.
      \end{center}
    \end{minipage}
  \end{center}
  \vspace*{\stretch{1}}
  \begin{abstract}
    \normalsize 
    In this paper, we present a formalism for computing the Yukawa
    couplings in heterotic standard models. This is accomplished by
    calculating the relevant triple products of cohomology groups,
    leading to terms proportional to $QHu$, $Q\bar{H}d$, $LH\nu$ and
    $L\bar{H}e$ in the low energy superpotential. These interactions
    are subject to two very restrictive selection rules arising from
    the geometry of the Calabi-Yau manifold. We apply our formalism to
    the ``minimal'' heterotic standard model whose observable sector
    matter spectrum is exactly that of the MSSM. The non-vanishing
    Yukawa interactions are explicitly computed in this context. These
    interactions exhibit a texture rendering one out of the three
    quark/lepton families naturally light.
  \end{abstract}
  \vspace*{\stretch{5}}
  \begin{minipage}{\textwidth}
    \underline{\hspace{5cm}}
    \centering
    \\
    Email: 
    \texttt{vbraun, ovrut@physics.upenn.edu, yang-hui.he@merton.ox.ac.uk}.
  \end{minipage}
\end{titlepage}

\tableofcontents

\section{Introduction}

Obtaining non-vanishing Yukawa couplings is one of the most important
issues in realistic superstring model building~\cite{Binetruy:2005ez}.
In this paper, we present a formalism for computing these terms and
explicitly demonstrate, within an important class of $E_8 \times E_8$
superstring vacua, that non-vanishing Yukawa couplings are generated
in the low energy effective theory.

In a series of papers~\cite{HetSM1, HetSM2, HetSM3}
and~\cite{MinimalHetSM}, we presented a class of ``heterotic standard
model'' vacua within the context of the $E_8 \times E_8$ heterotic
superstring. The observable sector of a heterotic standard model
vacuum is $N=1$ supersymmetric and consists of a stable, holomorphic
vector bundle, $V$, with structure group $SU(4)$ over an elliptically
fibered Calabi-Yau threefold, $X$, with a $\z3z3$ fundamental group.
In~\cite{HetSM1,HetSM2,HetSM3}, we gave non-trivial checks on the
slope-stability of the vector bundle $V$. A rigorous proof of the
stability of this bundle was presented in~\cite{Gomez:2005ii}. The
vector bundle $V$ in~\cite{MinimalHetSM} is also slope-stable. This
will be shown in detail in~\cite{future}. Each such bundle admits a
gauge connection which, in conjunction with a Wilson line,
spontaneously breaks the observable sector $E_8$ gauge group down to
the $SU(3)_{C} \times SU(2)_{L} \times U(1)_{Y}$ standard model group
times an additional gauged $U(1)_{B-L}$ symmetry. The spectrum arises
as the cohomology of the vector bundle $V$. For the vacuum presented
in~\cite{MinimalHetSM}, the matter spectrum is found to be precisely
that of the minimal supersymmetric standard model (MSSM). For this
reason, we refer to~\cite{MinimalHetSM} as the ``minimal'' heterotic
standard model. The vacua presented in~\cite{HetSM1, HetSM2,HetSM3}
also have the matter spectrum of the MSSM, with the exception of one
additional pair of Higgs--Higgs conjugate superfields. These vacua
contain \emph{no exotic multiplets} and \emph{no vector-like pairs of
  fields} with the exception of the Higgs pairs. They exist for both
weak and strong string coupling.  All previous attempts to find
realistic particle physics vacua in superstring
theories~\cite{Gross:1985rr, Sen:1985eb, Evans:1985vb, Breit:1985ud,
  Aspinwall:1987cn, Green:1987mn, Curio:1998vu, Andreas:1999ty,
  Donagi:1999gc, Krause:2000gp, Andreas:2003zb, Curio:2003ur,
  Curio:2004pf, Blumenhagen:2005ga, Blumenhagen:2005zg, Becker:2005sg,
  Raby:2005vc} have run into difficulties. These include predicting
extra vector-like pairs of light fields, multiplets with exotic
quantum numbers in the low energy spectrum, enhanced gauge symmetries
and so on. Heterotic standard models avoid all of these problems. As
for the hidden sector, there is no known obstruction to making it
$N=1$ supersymmetric as well, but we have not yet constructed the
requisite hidden sector bundle. It is also unclear whether that is
even phenomenologically desirable. In any case, in this paper we
consider only the visible sector interactions.

Elliptically fibered Calabi-Yau threefolds with $\Z_2$ and $\Z_2
\times \Z_2$ fundamental group were first constructed
in~\cite{SM-bundle1, SM-bundle2, SM-bundle3} and~\cite{z2z2-1,
  z2z2-2}, respectively.  More recently, the existence of elliptic
Calabi-Yau threefolds with $\z3z3$ fundamental group was demonstrated
and their classification given in~\cite{dP9Z3Z3}.
In~\cite{DonagiPrincipal, FMW1, FMW2, FMW3}, methods for building
stable, holomorphic vector bundles with arbitrary structure group in
$E_8$ over simply-connected elliptic Calabi-Yau threefolds were
introduced. These results were greatly expanded in a number of
papers~\cite{SM-bundle1, SM-bundle2, SM-bundle3, Diaconescu:1998kg,
  Donagi:2004qk, Donagi:2004ia} and then generalized to elliptically
fibered Calabi-Yau threefolds with non-trivial fundamental group
in~\cite{SM-bundle3, Ovrut:2002jk, z2z2-1, z2z2-2}. To obtain a
realistic spectrum, it was found necessary to introduce a new
method~\cite{SM-bundle1, SM-bundle2, SM-bundle3, z2z2-1, z2z2-2} for
constructing vector bundles. This method, which consists of building
the requisite bundles by ``extension'' from simpler, lower rank
bundles, was used for manifolds with ${\mathbb{Z}}_{2}$ fundamental
group in~\cite{MR1797016, MR1807601, SM-bundle3, SU5-z2-1, SU5-z2-2}
and in the heterotic standard model context in~\cite{dP9Z3Z3}.
In~\cite{HetSM1, HetSM2, HetSM3, SU5-z2-1, SU5-z2-2}, it was shown
that to compute the complete low-energy spectrum of such vacua one
must 1) evaluate the relevant sheaf cohomologies, 2) find the action
of the finite fundamental group on these spaces and, finally, 3)
tensor this with the action of the Wilson line on the associated
representation.  The low energy spectrum is the invariant cohomology
subspaces under the resulting group action. This method was applied
in~\cite{HetSM1, HetSM2, HetSM3, MinimalHetSM} to compute the exact
spectrum of all multiplets transforming non-trivially under the action
of the low energy gauge group. The accompanying natural method of
``doublet-triplet'' splitting was also discussed. A formalism was
presented in~\cite{ModHetSM} that allows one to enumerate and describe
the multiplets transforming trivially under the low energy gauge
group, namely, the vector bundle moduli.

Using the above, one can construct a class of heterotic standard
models and compute their entire low-energy spectrum. For example,
using a $\Z_2$ Wilson line one can break a $SU(5)$ GUT group to the
standard model gauge group.  Heterotic vacua in this context were
first computed in~\cite{SU5-z2-1, SU5-z2-2}. This was recently refined
in~\cite{Bouchard:2005ag} to construct a realistic heterotic standard
model with three chiral families of quarks/leptons and one pair of
Higgs--Higgs conjugate fields. One can also use orbifold CFT to arrive
at a minimal spectrum~\cite{Buchmuller:2005jr}. But for the purposes
of this paper we will be interested $\ZZZ$ Wilson lines breaking a
$Spin(10)$ GUT group to the standard model gauge group times
$U(1)_{B-L}$. As mentioned previously, the observable sector spectrum
consists exclusively of the three chiral families of quarks/leptons
(each family with a right-handed neutrino), either
one~\cite{MinimalHetSM} or two~\cite{HetSM1, HetSM2, HetSM3} pairs of
Higgs--Higgs conjugate fields and a small number of uncharged
geometric and vector bundle moduli. However, finding the particle
spectrum is far from the end of the story. To demonstrate that the
particle physics in these vacua is realistic, one must construct the
interactions of these fields in the low energy effective Lagrangian.
These interactions occur in two distinct parts of the action. Recall
that the matter part of an $N=1$ supersymmetric Lagrangian is
completely described in terms of two functions, the superpotential and
the K\"ahler potential. Of these, the superpotential, being a
``holomorphic'' function of chiral superfields, is much more amenable
to computation using methods of algebraic geometry. The superpotential
itself is a sum of several different pieces, such as Higgs $\mu$-terms
and Yukawa couplings. In a recent paper~\cite{HiggsMuTerm}, it was
shown how to compute Higgs $\mu$-terms in the superpotentials of
heterotic standard models.  In this paper, we continue our study of
holomorphic interactions by presenting a formalism for computing
Yukawa terms. We apply this method to calculate the Yukawa texture in
the minimal heterotic standard model~\cite{MinimalHetSM}.

Specifically, we do the following. In Section~\ref{sec:prelim}, we
review the relevant facts about the structure of heterotic standard
model vacua in general and the minimal heterotic vacuum in particular.
The formalism for computing the low energy spectrum is briefly
discussed and we give the results for the minimal heterotic standard
model vacuum.  The structure of Yukawa terms are then analyzed and
shown to occur as the product of three cohomology groups, two
corresponding to the quark/lepton doublets ($Q$,$L$) and singlets
($u$,$d$,$\nu$,$e$), and one corresponding to Higgs ($H$) and Higgs
conjugate ($\bar{H}$) fields in the effective low energy theory. It
follows that cubic terms of the form $QHu$, $Q\bar{H}d$, $LH\nu$ and
$L\bar{H}e$ are potentially generated in the superpotential.
Section~\ref{sec:LSS1} is devoted to discussing the first Leray
spectral sequence, which is associated with the projection of the
covering threefold $\tx$ onto the base space $B_2$. The Leray
decomposition of a sheaf cohomology group into $(p,q)$ subspaces is
discussed and applied to the cohomology spaces relevant to Yukawa
terms. It is shown that the triple product is subject to a $(p,q)$
selection rule which severely restricts the allowed non-vanishing
terms. The second Leray decomposition, associated with the projection
of the space $B_2$ onto its base $\IP1$, is presented in
Section~\ref{sec:LSS2}. The decomposition of any cohomology space into
its $[s,t]$ subspaces is discussed and applied to cohomologies
relevant to Yukawa terms. We show that Yukawa couplings are subject to
yet another selection rule associated with the $[s,t]$ decomposition.
Finally, it is demonstrated that the subspaces of cohomology that form
non-vanishing cubic terms project non-trivially onto both quark/lepton
doublets and singlets, as well as Higgs and Higgs conjugate fields
under the action of the $\z3z3$ group.

We conclude that non-vanishing Yukawa terms proportional to $QHu$,
$Q\bar{H}d$, $LH\nu$ and $L\bar{H}e$ appear in the low energy
superpotential of a minimal heterotic standard model. However, their
structure is constrained by the above selection rules. The exact
texture of the Yukawa interactions and its implications for the
quark/lepton mass matrix are presented in Section~\ref{sec:Yukawa}. We
show that, in a suitable basis, one out of the three quark/lepton
families is, prior to higher order and non-perturbative corrections,
massless. The remaining two generations have masses of the order of
the electroweak symmetry breaking scale.

\section{Preliminaries}
\label{sec:prelim}

\subsection{Heterotic String on a Calabi-Yau Manifold}
\label{sec:CY}

The observable sector of an $E_8 \times E_8$ heterotic standard model
vacuum consists of a stable, holomorphic vector bundle, $V$, over a
Calabi-Yau threefold, $X$. In particular, we are interested in an
$SU(4)$ instanton, breaking the low energy gauge group down to its
commutant
\begin{equation} 
  \xymatrix@C=15mm{
    E_8 \ar[r]^-{SU(4)} & Spin(10)
  }.
  \label{1}
\end{equation} 
Additionally, we want $\ZZZ$ Wilson lines $W$. The $Spin(10)$ group is
then spontaneously broken by the holonomy group of $W$ to
\begin{equation} 
  \xymatrix@C=15mm{
    Spin(10) 
    \ar[r]^-{\z3z3} &
    SU(3)_{C} \times SU(2)_{L} \times U(1)_{Y} \times U(1)_{B-L}
  }.
  \label{2}
\end{equation} 
In this way, the standard model gauge group emerges in the low energy
effective theory multiplied by an additional $U(1)$ gauge group whose
charges correspond to $B-L$ quantum numbers.

For $W$ to exist, the Calabi-Yau manifold $X$ must have fundamental
group $\ZZZ$. The physical properties of this vacuum are most easily
deduced not from $X$ and $V$ but, rather, from two closely related
entities, which we denote by $\Xt$ and $\Vt$ respectively. $\Xt$ is a
simply-connected Calabi-Yau threefold which admits a freely acting
$\z3z3$ group action such that
\begin{equation} 
  X= \Xt\Big/\big(\z3z3\big).
  \label{add1}
\end{equation}
That is, $\Xt$ is the universal covering space of $X$. Similarly,
$\Vt$ is a stable, holomorphic vector bundle over $\Xt$ with structure
group $SU(4)$ which is equivariant under the action of $\z3z3$. Then,
\begin{equation} 
  V=\Vt\Big/\big(\z3z3\big).
  \label{add2}
\end{equation}
The covering space $\tx$ for a heterotic standard model was discussed
in detail in~\cite{dP9Z3Z3}.  Here, it suffices to recall that $\tx$
is a fiber product
\begin{equation} 
  \tx=B_1 \times_{\IP1} B_2
  \label{3}
\end{equation}
of two rational elliptic ($\dP9$) surfaces $B_1$ and $B_2$ with $\ZZZ$
action. Thus, $\Xt$ is elliptically fibered over both surfaces with
the projections
\begin{equation} 
  \pi_1 : \tx \to B_1 \,, \quad \pi_2 : \tx \to B_2 \,.
  \label{4}
\end{equation}
The surfaces $B_1$ and $B_2$ are themselves elliptically fibered over
$\IP1$ with maps
\begin{equation}
  \label{5} 
  \beta_1 : B_1 \to \IP1 \,, \quad \beta_2 : B_2 \to \IP1 \,.
\end{equation}
Together, these projections yield the commutative diagram
\begin{equation}
  \label{6} 
  \vcenter{\xymatrix@!0@=12mm{ & \Xt \ar[dr]^{\pi_2}
      \ar[dl]_{\pi_1} \\ B_1 \ar[dr]_{\beta_1} & & B_2 \ar[dl]^{\beta_2} \\
      & \IP1 \,.  
    }}
\end{equation} 
The invariant homology ring of each special $\dP9$ surface is
generated by two $\z3z3$ invariant curve classes $f$ and $t$. Using
the projections in eq.~\eqref{4}, these can be pulled back to divisor
classes
\begin{equation} 
  \tau_1 = \pi_1^{-1}(t_{1}) 
  \,, \quad 
  \tau_2 = \pi_2^{-1}(t_{2}) 
  \,, \quad 
  \phi = \pi_1^{-1}(f_{1}) = \pi_2^{-1}(f_{2})
  \label{7}
\end{equation}
on $\Xt$. These three classes generate the even invariant homology
ring of $\Xt$. In particular,
\begin{equation}
  \Span\{\tau_{1},\tau_{2},\phi\} = 
  H^{2}\Big(\Xt,\mathbb{C}\Big)^{\ZZZ}
  \label{10}
\end{equation}
is the $\ZZZ$ invariant part of the K\"ahler moduli space.

\subsection{The Gauge Bundle}
\label{sec:bundle}

The crucial ingredient in a heterotic standard model is the choice of
the observable sector vector bundle $\Vt$. These bundles are
constructed using a generalization of the method of \emph{bundle
  extensions}~\cite{SM-bundle3,z2z2-2}.  Specifically, $\Vt$ is the
extension
\begin{equation}
  \label{8} 
  0 
  \longrightarrow
  V_1
  \longrightarrow
  \Vt
  \longrightarrow
  V_2
  \longrightarrow
  0
\end{equation}
of two rank two bundles $V_1$ and $V_2$ on $\Xt$. The solution for
$V_{1}$ and $V_{2}$ leading to the minimal heterotic standard model is
as follows. Define
\begin{equation}
  \label{9}
  \V1 =
  \oXt(-\tau_1+\tau_2) \otimes \pi_1^\ast(\W1)
  \,, \quad
  \V2 =
  \oXt(\tau_1-\tau_2) \otimes \pi_2^\ast(\W2)
  \,,
\end{equation}
where $\oXt(\mp\tau_1\pm\tau_2)$ are line bundles on $\Xt$ and the
rank $2$ bundles $\W1$, $\W2$ are constructed via an equivariant
version of the Serre construction as
\begin{equation}
  0 
  \longrightarrow
  \chi_1
  \oB1(- f_1) 
  \longrightarrow
  \W1
  \longrightarrow
  \chi_1^2
  \oB1( f_1) \otimes I_3^{B_1}
  \longrightarrow
  0
  \
  \label{11}
\end{equation}
and
\begin{equation}
  0 
  \longrightarrow
  \chi_2^2
  \oB2(-f_2) 
  \longrightarrow
  \W2
  \longrightarrow
  \chi_2
  \oB2(f_2) \otimes I_6^{B_2}
  \longrightarrow
  0    
  \,,
  \label{11A}
\end{equation}
where $I_3^{B_1}$ and $I_6^{B_2}$ denote the ideal sheaf\footnote{The
  analytic functions vanishing at the respective points.} of $3$ and
$6$ points in $B_1$ and $B_2$ respectively. The characters $\chi_1$
and $\chi_2$ are third roots of unity which generate the first and
second factors of $\ZZZ$. The observable sector equivariant bundle
$\Vt$ is then an invariant element of the space of extensions defined
in eq.~\eqref{8}. The vector bundle $\Vt$ so-constructed is
slope-stable~\cite{future}.

Let $R$ be any representation of $Spin(10)$ and $U(\Vt)_{R}$ the
associated tensor product bundle of $\Vt$.  Then, each sheaf
cohomology space $H^\ast\big(\Xt, U(\Vt)_{R}\big)$ carries a specific
representation of $\z3z3$. Similarly, the Wilson line $W$ manifests
itself as a $\z3z3$ group action on each representation $R$ of
$Spin(10)$.  As discussed in detail in~\cite{HetSM3}, the low-energy
particle spectrum is given by
\begin{multline}
  \label{12} 
  \ker\big(\dslash_{\Vt}\big) = \left( 
    H^0\big(\tx, \cO_{\tx}\big) \otimes \Rep{45} \right)^{\z3z3}
  \oplus \left( H^1\big(\tx, \tv^\dual \big) \otimes
    \barRep{16} \right)^{\z3z3} \\ 
  \oplus
  \left( H^1\big(\tx, \tv\big) \otimes
    \Rep{16} \right)^{\z3z3} \oplus 
  \left( H^1\big(\tx, \atv \big) \otimes \Rep{10} \right)^{\z3z3}
  \oplus 
  \left( H^1\big(\tx, \ad(\tv) \big) \otimes
    \Rep{1} \right)^{\z3z3}
  ,
\end{multline} 
where the superscript indicates the invariant subspace under the
action of $\z3z3$.  The invariant cohomology space $\big( H^0(\tx,
\cO_{\tx}) \otimes \Rep{45} \big)^{\z3z3}$ corresponds to gauge
superfields in the low-energy spectrum carrying the adjoint
representation of the gauge group. The matter cohomology spaces
\begin{equation}
  \Big( H^1(\tx, \tv^\dual ) \otimes \barRep{16}\Big)^{\z3z3}
  ,~
  \Big( H^1(\tx, \tv) \otimes \Rep{16} \Big)^{\z3z3}
  ,~
  \Big( H^1(\tx, \atv ) \otimes \Rep{10} \Big)^{\z3z3} 
\end{equation}
were all explicitly computed in~\cite{HetSM3}. One finds that
$H^{1}\big(\tx,\tv^\dual\big)=0$ and, hence, there are no vector-like
pairs of quark/lepton families.  The space $\big( H^1(\tx, \tv)
\otimes \Rep{16} \big)^{\z3z3}$ consists of three chiral families of
quarks/leptons, each family with a right-handed
neutrino~\cite{Giedt:2005vx}, and transforming as
\begin{equation}
  Q=\big(\Rep{3},   \Rep{2}, 1, 1 \big) \,,\quad
  u=\big(\barRep{3},\Rep{1}, -4, -1 \big) \,,\quad
  d=\big(\barRep{3},\Rep{1}, 2, -1 \big)
  \label{15A}
\end{equation}
and
\begin{equation}
  L=\big(\Rep{1},\Rep{2}, -3, -3 \big) \,,\quad
  e=\big(\Rep{1},\Rep{1}, 6, 3 \big) \,,\quad
  \nu=\big(\Rep{1},\Rep{1}, 0, 3 \big)
  \label{16A}
\end{equation}
under $SU(3)_{C} \times SU(2)_{L} \times U(1)_{Y} \times U(1)_{B-L}$.
We have displayed the quantum numbers $3Y$ and $3(B-L)$ for
convenience. The cohomology space $\big( H^1(\tx, \atv ) \otimes
\Rep{10} \big)^{\z3z3}$ is spanned by one vector-like pair of
Higgs--Higgs conjugate superfields
\begin{equation}
  H=\big( \Rep{1},\Rep{2}, 3, 0 \big) \,,\quad
  \bar{H}=\big( \Rep{1},\barRep{2}, -3,  0 \big)
  \,.
  \label{21A}
\end{equation}
That is, the matter spectrum is precisely that of the MSSM.  The
remaining cohomology space, $\big(H^1(\tx, \ad(\tv)) \otimes
\Rep{1}\big)^{\z3z3}$, was computed using the formalism introduced
in~\cite{ModHetSM} and corresponds to $13$ vector bundle moduli.

\subsection{Cubic Terms in the Superpotential}
\label{sec:couplings}

In this paper, we will focus on computing Yukawa terms.  It follows
from eq.~\eqref{12} that the $4$-dimensional Higgs and quark/lepton
fields correspond to certain $\bar{\partial}$-closed $(0,1)$-forms on
$\Xt$ with values in the vector bundle $\wedge^2 \Vt$ and $\Vt$
respectively. Since both $H$ and $\bar{H}$ arise from the same
cohomology space, we will denote either of these $1$-forms simply as
$\Psi^{H}$. For the same reason, we will schematically represent any
quark/lepton doublet by $\Psi^{(2)}$ and any singlet $1$-form by
$\Psi^{(1)}$, in any family. They can be written as
\begin{equation}
  \Psi^H = \psi^{H}_{\bar{\iota}[ab]}   
  \,\diff \bar{z}^{\bar{\iota}}
  , \quad
  \Psi^{(1)}= \psi_{\bar{\iota}a}^{(1)}
  \,\diff \bar{z}^{\bar{\iota}}
  , \quad
  \Psi^{(2)}= \psi_{\bar{\iota}b}^{(2)}
  \,\diff \bar{z}^{\bar{\iota}}
  , 
\end{equation}
where $a$, $b$ are valued in the $SU(4)$ bundle $\Vt$ and
$\{z^\iota,\bar{z}^{\bar{\iota}}\}$ are coordinates on the Calabi-Yau
threefold $\Xt$. Doing the dimensional reduction of the
$10$-dimensional Lagrangian yields cubic terms in the superpotential
of the $4$-dimensional effective action. It turns
out~\cite{Green:1987mn} that the coefficients of the cubic couplings
are simply the various allowed ways to obtain a number out of the
forms $\Psi^H$, $\Psi^{(1)}$, $\Psi^{(2)}$. That is
\begin{equation}
  W = \cdots + \lambda_{u}QHu + \lambda_{d}Q\bar{H}d
  +  \lambda_{\nu}LH\nu + \lambda_{e}L\bar{H}e 
\end{equation}
with the coefficients $\lambda$ determined by
\begin{equation}
  \label{eq:lambdaintegral}
  \begin{split}
    \lambda ~&= 
    \int_\Xt
    \Omega \wedge 
    \Tr\Big[ 
    \Psi^{(2)} \wedge \Psi^{H} \wedge \Psi^{(1)}
    \Big]
    = \\ &= 
    \int_\Xt
    \Omega \wedge 
    \Big( 
    \epsilon^{abcd}
    \psi^{(2)}_{\bar{\iota}a}
    \,
    \psi^{H}_{\bar{\kappa} [bc]} 
    \,
    \psi^{(1)}_{\bar{\epsilon}d}
    \Big)
    \diff \bar{z}^{\bar{\iota}} \wedge
    \diff \bar{z}^{\bar{\kappa}} \wedge
    \diff \bar{z}^{\bar{\epsilon}}
  \end{split}
\end{equation}
and $\Omega$ is the holomorphic $(3,0)$-form. Mathematically, we are
using the wedge product together with a contraction of the vector
bundle indices (that is, the determinant $\wedge^4\Vt = \oXt$) to
obtain a product
\begin{multline}
  \label{eq:cup}
  H^1\Big(\tx, \Vt \Big) \otimes 
  H^1\Big(\tx, \atv \Big) \otimes
  H^1\Big(\tx, \Vt \Big) 
  \longrightarrow \\ \longrightarrow
  H^3\Big(\tx, \Vt \otimes \atv \otimes \Vt\Big)
  \longrightarrow 
  H^3\Big(\tx, \cO_{\tx} \Big)  
\end{multline}
plus the fact that on the Calabi-Yau manifold $\Xt$
\begin{equation}
  H^3\Big(\Xt,\cO_{\Xt}\Big) =
  H^3\Big(\Xt, K_\Xt\Big) = 
  H_{\bar{\partial}}^{3,3}\Big(\Xt\Big) =
  H^6\Big(\Xt\Big)
\end{equation}
can be integrated over. If one were to use the heterotic string with
the ``standard embedding'', then the above product would simplify
further to the intersection of certain cycles in the Calabi-Yau
threefold~\cite{Greene:1987xh, Gepner:1988ck}. However, in our case
there is no such description.

Hence, to compute Yukawa terms, we must first analyze the cohomology
groups
\begin{equation} 
  H^1\Big(\tx, \Vt \Big),~
  H^1\Big(\tx, \atv \Big),~
  H^3\Big(\tx, \cO_{\tx} \Big)
  \label{14}
\end{equation}
and the action of $\z3z3$ on these spaces. We then have to evaluate
the product in eq.~\eqref{eq:cup}. As we will see in the following
sections, the two independent elliptic fibrations of $\Xt$ will force
some, but not all, products to vanish.

\section{The First Elliptic Fibration}
\label{sec:LSS1}

\subsection{The Leray Spectral Sequence}

As discussed in detail in~\cite{HetSM3}, the cohomology spaces on
$\tx$ are obtained by using two Leray spectral sequences.  In this
section, we consider the first of these sequences corresponding to the
projection
\begin{equation} 
  \tx \stackrel{\pi_{2}}{\longrightarrow} B_2.
  \label{15}
\end{equation}
For any sheaf $\cal{F}$ on $\tx$, the Leray spectral sequence tells us
that\footnote{In all the spectral sequences we are considering in this
  paper, higher differentials vanish trivially. Hence, the $E_2$ and
  $E_\infty$ tableaux are equal and we will not distinguish them in
  the following. Furthermore, there are no extension ambiguities for
  $\C$-vector spaces.}
\begin{equation} 
  H^i\Big( \tx, {\cal F} \Big) = \bigoplus^{p+q=i}_{p,q}
  H^p\Big(B_2, R^q\pi_{2\ast}\cF\Big),
  \label{16}
\end{equation}
where the only non-vanishing entries are for $p=0,1,2$ (since $\dim_\C
(B_2)=2$) and $q=0,1$ (since the fiber of $\tx$ is an elliptic curve,
therefore of complex dimension one). Note that the cohomologies
$H^{p}(B_2, R^q\pi_{2\ast}\cF)$ fill out the $2\times 3$
tableau\footnote{Recall that the zero-th derived push-down is just the
  ordinary push-down, $R^0\pi_{2\ast}=\pi_{2\ast}$.}
\begin{equation}
  \label{17} 
  \vcenter{ \def\w{34mm} \def\h{8mm} \xymatrix@C=0mm@R=0mm{
      {\scriptstyle q=1} & 
      *=<\w,\h>[F]{ H^0\big(B_2, R^1\pi_{2\ast}\cF\big) } &
      *=<\w,\h>[F]{ H^1\big(B_2, R^1\pi_{2\ast}\cF\big) } & 
      *=<\w,\h>[F]{ H^2\big(B_2, R^1\pi_{2\ast}\cF\big) } 
      \\ {\scriptstyle q=0} & 
      *=<\w,\h>[F]{ H^0\big(B_2, \pi_{2\ast}\cF\big) } & 
      *=<\w,\h>[F]{ H^1\big(B_2, \pi_{2\ast}\cF\big) } & 
      *=<\w,\h>[F]{ H^2\big(B_2, \pi_{2\ast}\cF\big) } 
      \\ & {\scriptstyle p=0} & {\scriptstyle p=1} & 
      {\scriptstyle p=2} 
    }} 
  \Rightarrow 
  H^{p+q}\Big(\Xt, \cF\Big)
  ,
\end{equation}
where ``$\Rightarrow H^{p+q}\big(\Xt, \cF\big)$'' reminds us of which
cohomology group the tableau is computing. Such tableaux are very
useful in keeping track of the elements of Leray spectral sequences.
As is clear from eq.~\eqref{16}, the sum over the diagonals yields the
desired cohomology of $\cF$. In the following, it will be very helpful
to define
\begin{equation} 
  H^p\Big(B_2, R^q\pi_{2\ast}\cF\Big) \equiv 
  \Hpq{\cF}{p}{q}
  .
  \label{18}
\end{equation}
Using this abbreviation, the tableau eq.~\eqref{17} reads
\begin{equation}
  \label{eq:Hpqtableau} 
  \vcenter{ \def\w{20mm} \def\h{8mm} \xymatrix@C=0mm@R=0mm{
      {\scriptstyle q=1} & 
      *=<\w,\h>[F]{ \Hpq{\cF}{0}{1} } &
      *=<\w,\h>[F]{ \Hpq{\cF}{1}{1} } &
      *=<\w,\h>[F]{ \Hpq{\cF}{2}{1} } 
      \\ {\scriptstyle q=0} & 
      *=<\w,\h>[F]{ \Hpq{\cF}{0}{0} } &
      *=<\w,\h>[F]{ \Hpq{\cF}{1}{0} } &
      *=<\w,\h>[F]{ \Hpq{\cF}{2}{0} } 
      \\ & {\scriptstyle p=0} & {\scriptstyle p=1} & 
      {\scriptstyle p=2} 
    }} 
  \Rightarrow 
  H^{p+q}\Big(\Xt, \cF\Big)
  .
\end{equation}

\subsection{Degrees and Products}


On the level of differential forms, we can understand the Leray
spectral sequence as decomposing differential forms into the number
$p$ of legs in the direction of the base and the number $q$ of legs in
the fiber direction. Obviously, this extra grading is preserved under
the wedge-product of the differential forms. Hence, any product
\begin{equation}
  H^i\Big( \Xt, \Fsheaf_1 \Big) \otimes
  H^j\Big( \Xt, \Fsheaf_2 \Big) 
  \longrightarrow
  H^{i+j}\Big( \Xt, \Fsheaf_1 \otimes \Fsheaf_2 \Big) 
\end{equation}
not only has to end up in overall degree $i+j$, but also has to
preserve the $(p,q)$-grading. That is,
\begin{equation}
  \label{eq:simpleproduct}
  \vcenter{\xymatrix@R=3ex{
      \Hpq{\Fsheaf_1}{p_1}{q_1}
      \ar@{}[r]|{\displaystyle \otimes} & 
      \Hpq{\Fsheaf_2}{p_2}{q_2}
      \ar[r] &
      \Hpq{\Fsheaf_1\otimes\Fsheaf_2}{p_1+p_2}{q_1+q_2}
      \\
      H^{p_1+q_1}\Big( \Xt, \Fsheaf_1 \Big) 
      \ar@{}[u]|{\displaystyle \cap}
      \ar@{}[r]|{\displaystyle \otimes} & 
      H^{p_2+q_2}\Big( \Xt, \Fsheaf_2 \Big) 
      \ar@{}[u]|{\displaystyle \cap}
      \ar[r] &
      H^{p_1+p_2+q_1+q_2}\Big( \Xt, \Fsheaf_1 \otimes \Fsheaf_2 \Big) 
      \ar@{}[u]|{\displaystyle \cap}
      \,.
    }}
\end{equation}
This is all we are going to need in the following, but we would like
to mention the following caveat. Although it does not happen here,
sometimes the push-down is not a vector bundle, but a (non-locally
free) sheaf. Then the identification with bundle-valued differential
forms is not possible. The way around this is well-known; one has to
replace the coherent sheaf by a complex of vector bundles. Now one can
again think in terms of differential forms, but at the cost of working
in the derived category. What can and does happen in general is the
appearance of derived tensor products. That is, the tensor product of
complexes may no longer be quasi-isomorphic to a complex with only one
entry. The effect is that the product ends up in
\begin{equation}
  \label{eq:derivedproduct}
  \Hpq{\Fsheaf_1}{p_1}{q_1}
  \otimes
  \Hpq{\Fsheaf_2}{p_2}{q_2}
  \longrightarrow
  \bigoplus_{n=0}^{\min(\hd(\Fsheaf_1),\hd(\Fsheaf_2))}
  \Hpq{\Fsheaf_1\otimes\Fsheaf_2}{p_1+p_2+n}{q_1+q_2-n}  
  \,,
\end{equation}
where $\hd\big(\Fsheaf_i\big)+1$ is the length of the shortest locally
free resolution of $\Fsheaf_i$. In all products that occur in this
paper $\hd(\Fsheaf)=0$ and, hence, eq.~\eqref{eq:derivedproduct}
simplifies to eq.~\eqref{eq:simpleproduct}.

\subsection{The First Leray Decomposition of the Volume Form}

Let us first discuss the $(p,q)$ Leray tableau for the sheaf
$\cF=\cO_{\tx}$, which is the last term in eq.~\eqref{14}. Since this
is the trivial line bundle, it immediately follows that
\begin{equation}
  \label{19} 
  \vcenter{ \def\w{20mm} \def\h{6mm} \xymatrix@C=0mm@R=0mm{
      {\scriptstyle q=1} & 
      *=<\w,\h>[F]{ 0 } & *=<\w,\h>[F]{ 0 } &
      *=<\w,\h>[F]{ \Rep{1} } 
      \\ {\scriptstyle q=0} & 
      *=<\w,\h>[F]{ \Rep{1} } & 
      *=<\w,\h>[F]{ 0 } & 
      *=<\w,\h>[F]{ 0 } 
      \\ & 
      {\scriptstyle p=0} & {\scriptstyle p=1} & {\scriptstyle p=2} 
    }} 
  \Rightarrow
  H^{p+q}\Big(\tx,\cO_\tx\Big)
  .
\end{equation}
From eqns.~\eqref{16} and~\eqref{19} we see that
\begin{equation} 
  H^3 \Big( \tx, \cO_\tx \Big) = \Hpq{\cO_\tx}{2}{1} = \Rep{1}
  ,
  \label{20}
\end{equation}
where the $\Rep{1}$ indicates that $H^3 ( \tx, \cO_{\tx} )$ is a
one-dimensional space carrying the trivial action of $\z3z3$.

\subsection{The First Leray Decomposition of Higgs Fields}

Now consider the $(p,q)$ Leray tableau for the sheaf $\cF = \atv$,
which is the second term in eq.~\eqref{14}. This can be explicitly
computed and is given by
\begin{equation}
  \label{22} 
  \vcenter{ \def\w{20mm} \def\h{6mm} \xymatrix@C=0mm@R=0mm{
      {\scriptstyle q=1} & 
      *=<\w,\h>[F]{ 0 } & 
      *=<\w,\h>[F]{ \rho_{4} } & 
      *=<\w,\h>[F]{ 0 } 
      \\ {\scriptstyle q=0} & 
      *=<\w,\h>[F]{ 0 } &
      *=<\w,\h>[F]{ \rho_{4} } & 
      *=<\w,\h>[F]{ 0 } \\ &
      {\scriptstyle p=0} & {\scriptstyle p=1} & {\scriptstyle p=2} 
    }} 
  \Rightarrow 
  H^{p+q}\Big(\Xt, \atv\Big)
  ,
\end{equation}
where $\rho_{4}$ is the four-dimensional representation
\begin{equation} 
  \rho_{4}= 
  \chi_{2} \oplus \chi_{2}^{2} \oplus \chi_{1}
  \chi_{2}^{2} \oplus \chi_{1}^{2} \chi_{2} 
  \label{23}
\end{equation}
of $\z3z3$. In general, it follows from eq.~\eqref{16} that $H^1(\tx,
\atv )$ is the sum of the two subspaces $\Hpq{\atv}{0}{1} \oplus
\Hpq{\atv}{1}{0}$.  However, we see from the Leray tableau
eq.~\eqref{22} that the $\Hpq{\atv}{0}{1}$ space vanishes. Hence,
\begin{equation} 
  H^1\Big(\tx, \atv \Big) = 
  \Hpq{\atv}{1}{0} =
  \rho_{4}  
  .
  \label{24}
\end{equation}

\subsection{The First Leray Decomposition of the Quark/Lepton Fields}

Now consider the $(p,q)$ Leray tableau for the sheaf $\cF = \Vt$,
which is the first term in eq.~\eqref{14}. This can be explicitly
computed and is given by
\begin{equation}
  \label{26a} 
  \vcenter{ \def\w{20mm} \def\h{6mm} \xymatrix@C=0mm@R=0mm{
      {\scriptstyle q=1} & 
      *=<\w,\h>[F]{ RG } & 
      *=<\w,\h>[F]{ 0 } & 
      *=<\w,\h>[F]{ 0 } 
      \\ {\scriptstyle q=0} & 
      *=<\w,\h>[F]{ 0 } &
      *=<\w,\h>[F]{ RG^{\oplus2} } & 
      *=<\w,\h>[F]{ 0 } \\ &
      {\scriptstyle p=0} & {\scriptstyle p=1} & {\scriptstyle p=2} 
    }} 
  \Rightarrow 
  H^{p+q}\Big(\Xt, \Vt\Big)
  ,
\end{equation}
where $RG$ is the regular representation of $\z3z3$ given by
\begin{equation} 
  RG=
  1 \oplus \chi_{1} \oplus \chi_{2} \oplus \chi_{1}^{2} \oplus 
  \chi_{2}^{2} \oplus \chi_{1}\chi_{2} \oplus \chi_{1}
  \chi_{2}^{2} \oplus \chi_{1}^{2} \chi_{2} \oplus \chi_{1}^{2} \chi_{2}^{2}. 
  \label{27a}
\end{equation}
It follows from eq.~\eqref{16} that $H^1(\tx, \Vt )$ is the sum of the
two subspaces
\begin{equation} 
  H^1\Big(\tx, \Vt \Big) = \Hpq{\Vt}{0}{1} \oplus  \Hpq{\Vt}{1}{0}
  .
  \label{28a}
\end{equation}
Furthermore, eq.~\eqref{26a} tells us that
\begin{equation}
  \Hpq{\Vt}{0}{1}= RG, \quad \Hpq{\Vt}{1}{0} = RG^{\oplus2}.
  \label{29a}
\end{equation}
Technically, the structure of eq.~\eqref{28a} is associated with the
fact that the cohomology $H^{\ast}\big(\Xt,\Vt\big)$ decomposes into
$H^{\ast}\big(\Xt,V_{1}\big) \oplus H^{\ast}\big(\Xt,V_{2}\big)$. It
turns out that the two subspaces in eq.~\eqref{28a} arise as
\begin{equation}
  RG=H^{1}\big(\Xt,V_{1}\big), \quad RG^{\oplus2}=H^{1}\big(\Xt,V_{2}\big)
  \label{hope4}
\end{equation}
respectively.

\subsection{The (p,q) Selection Rule}

Having computed the decompositions of $H^3(\tx, \cO_{\tx} )$,
$H^1(\tx, \atv)$ and $H^1(\tx, \Vt )$ into their ${(p,q)}$ Leray
subspaces, we can now analyze the $(p,q)$ components of the triple
product
\begin{equation}
  \label{eq:product}
  H^1\Big(\tx, \Vt \Big) \otimes 
  H^1\Big(\tx, \atv \Big) \otimes
  H^1\Big(\tx, \Vt \Big) 
  \longrightarrow 
  H^3\Big(\tx, \cO_{\tx} \Big)  
\end{equation}
given in eq.~\eqref{eq:cup}. Inserting eqns.~\eqref{24}
and~\eqref{28a}, we see that
\begin{multline}
  \label{eq:pqCup}
  H^1\Big(\tx, \Vt \Big) \otimes 
  H^1\Big(\tx, \atv \Big) \otimes
  H^1\Big(\tx, \Vt \Big) 
  = \\ 
  \Big( \Hpq{\Vt}{0}{1} \oplus 
  \Hpq{\Vt}{1}{0} \Big)
  \otimes
  \Hpq{\atv}{1}{0} 
  \otimes 
  \Big( \Hpq{\Vt}{0}{1} \oplus 
  \Hpq{\Vt}{1}{0} \Big)
  = \\ 
  \underbrace{\Big( 
    {\scriptstyle 
      \Hpq{\Vt}{0}{1} \otimes
      \Hpq{\atv}{1}{0} \otimes 
      \Hpq{\Vt}{1}{0} 
    }
    \Big)^{\oplus2} 
  }_{\text{total $(p,q)$ degree }=\, (2,1)}
  \oplus
  \underbrace{\Big( 
    {\scriptstyle 
      \Hpq{\Vt}{1}{0} \otimes
      \Hpq{\atv}{1}{0} \otimes 
      \Hpq{\Vt}{1}{0} 
    }
    \Big) 
  }_{\text{total $(p,q)$ degree }=\, (3,0)}
  \oplus
  \underbrace{\Big( 
    {\scriptstyle 
      \Hpq{\Vt}{0}{1} \otimes
      \Hpq{\atv}{0}{1} \otimes 
      \Hpq{\Vt}{0}{1} 
    }
    \Big) 
  }_{\text{total $(p,q)$ degree }=\, (0,3)}
\end{multline}
Because of the $(p,q)$ degree, we see from eq.~\eqref{20} that only
the first term can have a non-zero product in
\begin{equation}
  \label{48}
  H^3 \Big( \tx, \cO_\tx \Big) = \Hpq{\cO_\tx}{2}{1}
  .
\end{equation}
It follows that the first quark/lepton family, which arises from
\begin{equation} 
  \Hpq{\Vt}{0}{1} = RG,
  \label{49}
\end{equation}
will form non-vanishing Yukawa terms with the second and third
quark/lepton families coming from
\begin{equation} 
  \Hpq{\Vt}{1}{0} = RG^{\oplus2}.
  \label{49a}
\end{equation}
All other Yukawa couplings must vanish. We refer to this as the
\emph{$(p,q)$ Leray degree selection rule}. We conclude that the only
non-zero product in eq.~\eqref{eq:product} is of the form
\begin{equation}
  \label{eq:pqproduct}
  \Hpq{\Vt}{0}{1} \otimes
  \Hpq{\atv}{1}{0} \otimes 
  \Hpq{\Vt}{1}{0} 
  \longrightarrow
  \Hpq{\cO_\Xt}{2}{1}
  .
\end{equation}
Roughly what happens is the following. The holomorphic $(3,0)$-form
$\Omega$ has two legs in the base and one leg in the fiber direction.
According to eq.~\eqref{24}, both $1$-forms $\Psi^H$ corresponding to
Higgs and Higgs conjugate have their one leg in the base direction.
Therefore, the wedge product in eq.~\eqref{eq:lambdaintegral} can only
be non-zero if one quark/lepton $1$-form $\Psi$ has its leg in the
base direction and the other quark/lepton $1$-form $\Psi$ has its leg
in the fiber direction.

We conclude that due to a selection rule for the $(p,q)$ Leray degree,
the Yukawa terms in the effective low energy theory can involve only a
coupling of the first quark/lepton family to the second and third.
All other Yukawa couplings must vanish.

\section{The Second Elliptic Fibration}
\label{sec:LSS2}

\subsection{The Second Leray Spectral Sequence}

So far, we only made use of the fact that our Calabi-Yau manifold is
an elliptic fibration over the base $B_2$. But the $\dP9$ surface
$B_2$ is itself elliptically fibered over $\IP1$. Consequently,
there is yet another selection rule coming from the second elliptic
fibration. Therefore, we now consider the second Leray spectral
sequence corresponding to the projection
\begin{equation} 
  B_2 \stackrel{\beta_{2}}{\longrightarrow}
  \IP1.
  \label{50}
\end{equation}
For any sheaf $\cFt$ on $B_2$, the Leray sequence now starts with a $2
\times 2$ Leray tableau
\begin{equation}
  \label{52} 
  \vcenter{ \def\w{40mm} \def\h{8mm} \xymatrix@C=0mm@R=0mm{
      {\scriptstyle t=1} & 
      *=<\w,\h>[F]{ H^0\big( \IP1, R^1\beta_{2\ast} \cFt\big) } & 
      *=<\w,\h>[F]{ H^1\big( \IP1, R^1\beta_{2\ast}\cFt\big) } \\ 
      {\scriptstyle t=0} &
      *=<\w,\h>[F]{ H^0\big( \IP1, \beta_{2\ast} \cFt \big) } &
      *=<\w,\h>[F]{ H^1\big( \IP1, \beta_{2\ast} \cFt \big) } \\ &
      {\scriptstyle s=0} & {\scriptstyle s=1} }} 
  \Rightarrow
  H^{s+t}\Big(B_2, \cFt \Big)
  .
\end{equation}
Again, the sum over the diagonals yields the desired cohomology of
$\cFt$. Note that to evaluate the product eq.~\eqref{eq:pqproduct}, we
need the $[s,t]$ Leray tableaux for
\begin{equation}
  \cFt = 
  R^1\pi_{2\ast} \big(\Vt \big)
  ,~
  \pi_{2\ast} \big(\Vt \big)
  ,~
  \pi_{2\ast} \big( \atv \big)
  ,~
  R^1\pi_{2\ast} \big(\cO_\Xt\big)
  .
\end{equation}
In the following, it will be useful to define
\begin{equation} 
  H^{s} \bigg( \IP1, 
  R^t \beta_{2*} \Big( R^q \pi_{2\ast}\big(\cF\big)  \Big) \bigg) 
  \equiv 
  \Hst{\cF}{q}{s}{t}
  .
  \label{53}
\end{equation}
One can think of $\Hst{\cF}{q}{s}{t}$ as the subspace of
$H^\ast\big(\Xt, \cF\big)$ that can be written as forms with $q$ legs
in the $\pi_2$-fiber direction, $t$ legs in the $\beta_2$-fiber
direction, and $s$ legs in the base $\IP1$ direction.

\subsection{The Second Leray Decomposition of the Volume Form}

Let us first discuss the $[s,t]$ Leray tableau for
$\cFt=R^1\pi_{2\ast} \big(\cO_{\tx}\big)=K_{B_2}$, the canonical line
bundle. It follows immediately that
\begin{equation}
  \label{54} 
  \vcenter{ \def\w{20mm} \def\h{6mm} \xymatrix@C=0mm@R=0mm{
      {\scriptstyle t=1} & *=<\w,\h>[F]{ 0 } & *=<\w,\h>[F]{ \Rep{1} }
      \\ {\scriptstyle t=0} & *=<\w,\h>[F]{ 0 } & *=<\w,\h>[F]{ 0 } \\ &
      {\scriptstyle s=0} & {\scriptstyle s=1} }} 
  \Rightarrow
  H^{s+t}\Big(B_2, R^1\pi_{2\ast} \big(\cO_{\tx}\big) \Big)  
  .
\end{equation}
In our notation, this means that
\begin{equation}
  H^2\Big(B_2, R^1\pi_{2\ast} \big(\cO_{\tx}\big)\Big) = 
  \Hst{\cO_\Xt}{1}{1}{1}
\end{equation}
has pure $[s,t]=[1,1]$ degree. To summarize, we see that
\begin{equation} 
  \label{55}
  H^3\Big( \Xt, \cO_\Xt \Big) = 
  \Hpq{\cO_\Xt}{2}{1} =
  \Hst{\cO_\Xt}{1}{1}{1} =
  \Rep{1}
  .
\end{equation}

\subsection{The Second Leray Decomposition of Higgs Fields}

Now consider the $[s,t]$ Leray tableau for the sheaf
$\cFt=\pi_{2\ast}\big(\atv\big)$.  This can be explicitly computed 
and is given by
\begin{equation}
  \label{57} 
  \vcenter{ \def\w{35mm} \def\h{8mm} \xymatrix@C=0mm@R=0mm{
      {\scriptstyle t=1} & 
      *=<\w,\h>[F]{ 
        \chi_{1}\chi_{2}^{2} 
      } & 
      *=<\w,\h>[F]{ 0 } \\
      {\scriptstyle t=0} & 
      *=<\w,\h>[F]{ 0 } & 
      *=<\w,\h>[F]{ 
        \chi_{2} \oplus \chi_{2}^{2} \oplus \chi_{1}^{2}\chi_{2}}  
      \\ & {\scriptstyle s=0} &
      {\scriptstyle s=1} }} 
  \Rightarrow 
  H^{s+t}\Big(B_2, \pi_{2\ast} \big(\atv\big) \Big)    
  .
\end{equation}
This means that the $4$ copies of the $\Rep{10}$ of $Spin(10)$ given
in eq.~\eqref{24} split as
\begin{equation}
  H^1\Big( \Xt, \atv \Big) = 
  \Hpq{\atv}{1}{0} = 
  \Hst{\atv}{0}{0}{1} \oplus \Hst{\atv}{0}{1}{0},
  \label{beta}
\end{equation}
where
\begin{equation}
  \label{58}
  \begin{split}
    \Hst{\atv}{0}{0}{1} ~&=
    \chi_{1}\chi_{2}^{2}
    \\
    \Hst{\atv}{0}{1}{0} ~&=    
    \chi_{2} \oplus \chi_{2}^{2} \oplus \chi_{1}^{2}\chi_{2}.
  \end{split}
\end{equation}
Note that
\begin{equation} 
  \label{60}
  \Hst{\atv}{0}{0}{1} \oplus \Hst{\atv}{0}{1}{0} = \rho_{4}
\end{equation}
in eq.~\eqref{23}, as it must.

\subsection{The Second Leray Decomposition of the Quark/Lepton Fields}

Finally, let us consider the $[s,t]$ Leray tableau for the
quark/lepton fields. We have already seen that, due to the $(p,q)$
selection rule, both the first quark/lepton family arising from
\begin{equation}  
  \Hpq{\Vt}{0}{1} = RG
  \label{AA}
\end{equation}
and the second and third quark/lepton families coming from
\begin{equation}   
  \Hpq{\Vt}{1}{0} = RG^{\oplus2}
  \label{BB}
\end{equation}
must occur in non-vanishing Yukawa interactions.  Therefore, we are
only interested in the $[s,t]$ decomposition of each of these
subspaces. The $ \Hpq{\Vt}{0}{1}$ subspace is associated with the
degree $0$ cohomology of the sheaf $R^1\pi_{2\ast} \big( \Vt \big)$.
The corresponding Leray tableau is given by
\begin{equation}
  \label{61} 
  \vcenter{ \def\w{20mm} \def\h{6mm} \xymatrix@C=0mm@R=0mm{
      {\scriptstyle t=1} & 
      *=<\w,\h>[F]{ 0 } & 
      *=<\w,\h>[F]{ 0 } \\
      {\scriptstyle t=0} & 
      *=<\w,\h>[F]{ RG } & 
      *=<\w,\h>[F]{ 0 }
      \\ & {\scriptstyle s=0} & {\scriptstyle s=1} }} 
  \Rightarrow
  H^{s+t}\Big(B_2, R^1\pi_{2\ast} \big(\Vt \big)
  \Big)     
  .
\end{equation}
It follows that the first family of quarks/leptons has $[s,t]$ degree
$[0,0]$,
\begin{equation} 
  \label{62}
  \Hpq{\Vt}{0}{1} = 
  \Hst{\Vt}{1}{0}{0}= 
  RG
  .
\end{equation}
The $ \Hpq{\Vt}{1}{0}$ subspace is associated with the degree $1$
cohomology of the sheaf $\pi_{2\ast} \big( \Vt \big)$. The
corresponding Leray tableau is given by
\begin{equation}
  \label{61a} 
  \vcenter{ \def\w{20mm} \def\h{6mm} \xymatrix@C=0mm@R=0mm{
      {\scriptstyle t=1} & 
      *=<\w,\h>[F]{ RG^{\oplus2} } & 
      *=<\w,\h>[F]{ 0 } \\
      {\scriptstyle t=0} & 
      *=<\w,\h>[F]{ 0 } & 
      *=<\w,\h>[F]{ 0 }
      \\ & {\scriptstyle s=0} & {\scriptstyle s=1} }} 
  \Rightarrow
  H^{s+t}\Big(B_2,\pi_{2\ast} \big(\Vt \big)
  \Big)     
  .
\end{equation}
It follows that the second and third families of quarks/leptons has
$[s,t]$ degree $[0,1]$,
\begin{equation} 
  \label{62a}
  \Hpq{\Vt}{1}{0} = 
  \Hst{\Vt}{1}{0}{1}= 
  RG^{\oplus2}
  .
\end{equation}

\subsection{The [s,t] Selection Rule}

Having computed the decompositions of the relevant cohomology spaces
into their $[s,t]$ Leray subspaces, we can now calculate the triple
product eq.~\eqref{eq:cup}. The $(p,q)$ selection rule dictates that
the only non-zero product is of the form eq.~\eqref{eq:pqproduct}. Now
split each term in this product into its $[s,t]$ subspaces, as given
in eqns.~\eqref{55},~\eqref{58}, and~\eqref{62} respectively. The
result is
\begin{multline} 
  \label{65}
  \Hst{\Vt}{1}{0}{0} 
  \otimes
  \Big( \Hst{\atv}{0}{0}{1} \oplus \Hst{\atv}{0}{1}{0} \Big)
  \otimes 
  \Hst{\Vt}{1}{0}{1}
  \longrightarrow
  \Hst{\cO_\Xt}{1}{1}{1} 
  .
\end{multline}
Clearly, this triple product vanishes by degree unless we choose the
$\Hst{\atv}{0}{1}{0}$ from the $\Hpq{\atv}{1}{0}$ subspace.  In this
case, eq.~\eqref{65} becomes
\begin{equation} 
  \label{65-1}
  \Hst{\Vt}{1}{0}{0} 
  \otimes
  \Hst{\atv}{0}{1}{0}
  \otimes
  \Hst{\Vt}{1}{0}{1}
  \longrightarrow
  \Hst{\cO_\Xt}{1}{1}{1} 
  ,
\end{equation}
which is consistent.

We conclude that there is, in addition to the $(p,q)$ selection rule
discussed above, a \emph{$[s,t]$ Leray degree selection rule}. This
rule continues to allow non-vanishing Yukawa couplings of the first
quark/lepton family with the second and third quark/lepton families,
but only through the
\begin{equation}
  \Hst{\atv}{0}{1}{0}= \chi_{2} \oplus \chi_{2}^{2} \oplus \chi_{1}^{2}\chi_{2}
  \label{alpha}
\end{equation}
component of $\Hpq{\atv}{1}{0}$ in eq.~\eqref{beta}.

\subsection{Wilson Lines}

We have, in addition to the $SU(4)$ instanton, a non-vanishing Wilson
line. Its effect is to break the $Spin(10)$ gauge group down to the
desired $SU(3)_{C} \times SU(2)_{L} \times U(1)_{Y} \times U(1)_{B-L}$
gauge group.  First, consider the $\Rep{16}$ matter representations.
We choose the Wilson line $W$ so that its $\z3z3$ action on each
$\Rep{16}$ is given by
\begin{equation}
  \Rep{16}= \Big[
  \chi_{1}\chi_{2}^{2} Q
  \oplus \chi_{2}^{2} e
  \oplus \chi_{1}^{2}\chi_{2}^{2} u \Big]
  \oplus 
  \Big[ L
  \oplus \chi_{1}^{2} d \Big]
  \oplus \chi_{2} \nu,
  \label{hope1}
\end{equation}
where the representations $Q$,$u$,$d$ and $L$,$\nu$,$e$ were defined
in eqns.~\eqref{15A} and~\eqref{16A}, respectively. Recall from
eqns.~\eqref{28a} and~\eqref{29a} that $H^{1}\big( \Xt, \Vt \big)=RG
\oplus RG^{\oplus2}$.  Tensoring any $RG$ subspace of the cohomology
space $H^{1}\big(\Xt,\Vt\big)$ with a $\Rep{16}$ using
eqns.~\eqref{27a} and~\eqref{hope1}, we find that the invariant
subspace under the $\z3z3$ action is
\begin{equation}
  \Big( RG \otimes \Rep{16} \Big)^{\z3z3}= \Span \big\{Q,u,d,L,\nu,e \big\}
  \label{hope2}
\end{equation}
It follows that each $RG$ subspace of $H^{1}\big(\Xt,\Vt\big)$
projects to a complete quark/lepton family at low energy. This
justifies our identification of the subspace $RG$ with the first
quark/lepton family and the subspace $RG^{\oplus2}$ with the second
and third quark/lepton families throughout the text.

Second, notice that each fundamental matter field in the $\Rep{10}$
can be broken to a Higgs field, a color triplet, or projected out. In
particular, we are going to choose the Wilson line $W$ so that its
$\z3z3$ action on a $\Rep{10}$ representation of $Spin(10)$ is given
by
\begin{equation} 
  \Rep{10}=
  \Big[ 
  \chi_{2}^{2} H \oplus \chi_{1}^{2} \chi_{2}^{2} C
  \Big] \oplus \Big[ 
  \chi_{2} \bar{H} \oplus \chi_{1} \chi_{2}
  \bar{C}
  \Big]
  ,
  \label{66}
\end{equation}
where $H$ and $\bar{H}$ are defined in eq.~\eqref{21A} and
\begin{equation}  
  C=\big( \Rep{3},\Rep{1}, -2, -2 \big),\quad 
  \bar{C}=\big( \barRep{3}, \Rep{1}, 2, 2 \big)
  \label{67}
\end{equation}
are the color triplet representations of $SU(3)_{C} \times SU(2)_{L}
\times U(1)_{Y} \times U(1)_{B-L}$.  Tensoring this with the
cohomology space $H^1\big(\Xt,\atv\big)$, we find the invariant
subspace under the combined $\z3z3$ action on the cohomology space and
the Wilson line to be
\begin{equation}
  \Big( H^1\big(\Xt,\atv\big) \otimes \Rep{10} \Big)^\z3z3
  = 
  \Span
  \big\{H,\bar{H}\big\}
  .
\end{equation}
Hence, precisely one pair of Higgs--Higgs conjugate fields survives the
$\z3z3$ quotient. As required for any realistic model, all color
triplets are projected out.  The new information now is the $(p,q)$
and $[s,t]$ degrees of the Higgs fields. Using the decompositions
eqns.~\eqref{24} and~\eqref{beta} of $H^1\big(\Xt,\atv\big)$, we find
\begin{multline}
  \Big( H^1\big(\Xt,\atv\big) \otimes \Rep{10} \Big)^\z3z3
  =
  \Big( \Hpq{\atv}{1}{0} \otimes \Rep{10} \Big)^\z3z3
  = \\ =
  \underbrace{
    \Big( \Hst{\atv}{0}{0}{1} \otimes \Rep{10} \Big)^\z3z3
  }_{= 0 }
  \oplus 
  \underbrace{
    \Big( \Hst{\atv}{0}{1}{0} \otimes \Rep{10} \Big)^\z3z3
  }_{= \Span\{H, \bar{H}\} }
  .
\end{multline}
The dimensions and basis' of the two terms on the right side of this
expression are determined by taking the tensor product of
eqns.~\eqref{58} and~\eqref{66} and keeping the $\z3z3$ invariant
part. Note that the subspace forming the non-zero Yukawa couplings in
eq.~\eqref{65-1}, namely $\Hst{\atv}{0}{1}{0}$, indeed projects to the
Higgs--Higgs conjugate pair in the low energy theory.

\section{Yukawa Couplings}
\label{sec:Yukawa}

To conclude, we analyzed cubic terms in the superpotential of the form
\begin{equation} 
  \lambda_{u,ij} Q_{i} H u_{j},  \quad 
  \lambda_{d,ij} Q_{i} \bar{H} d_{j}, \quad 
  \lambda_{\nu,ij} L_{i} H \nu_{j},  \quad 
  \lambda_{e,ij} L_{i} \bar{H} e_{j} 
  \label{67-1}
\end{equation}
where
\begin{itemize}
\item each coefficient $\lambda$ is determined by an integral
  of the form of eq.~\eqref{eq:lambdaintegral},
\item $Q_i$,$L_{i}$ for $i=1,2,3$ are the electroweak doublets of the three
  quarks/lepton families respectively,
\item $u_{j}$,$d_{j}$,$\nu_{j}$,$e_{j}$ for $j=1,2,3$ are the
  electroweak singlets of the three quark/lepton families
  respectively,
\item $H$ is the Higgs field, and
\item $\bar{H}$ is the Higgs conjugate field.
\end{itemize}
We found that they are subject to two independent selection rules
coming from the two independent torus fibrations. The first selection
rule is that the total $(p,q)$ degree is $(2,1)$.  Since the $(p,q)$
degrees for the first quark/lepton family, the second and third
quark/lepton families and the Higgs fields are $(0,1)$, $(1,0)$ and
$(1,0)$ respectively, it follows that the only non-vanishing $\lambda$
coefficients are of the form
\begin{equation}
  \lambda_{u,1j}, \lambda_{u,j1} \quad \lambda_{d,1j}, \lambda_{d,j1}\quad 
  \lambda_{\nu,1j}, \lambda_{\nu,j1}
  \quad \lambda_{e,1j}, \lambda_{e,j1}
  \label{gamma}
\end{equation}
for $j=2,3$. That is, the only non-zero Yukawa terms couple the first
family to the second and third families respectively.  The second
selection rule imposes independent constraints. It states that the
total $[s,t]$ degree has to be $[1,1]$. Of the two possible $[s,t]$
degrees associated with the Higgs fields, only the $[1,0]$ subspace
satisfies the $[s,t]$ selection rule. Happily, this is precisely the
component that projects to a $H$-$\bar{H}$ pair at low energy. Hence,
the conclusion in eq.~\eqref{gamma} is unaltered.

Let us analyze, for example, the Yukawa contribution to the up-quark
mass matrix. Assuming that $H$ gets a non-vanishing vacuum expectation
value $\langle H \rangle$ in its charge neutral component, this
contribution can be written as
\begin{equation}
  \left(
    \begin{array}{ccc}
      0 & \lambda_{u,12}\langle H \rangle & \lambda_{u,13}\langle H \rangle \\
      \lambda_{u,21}\langle H \rangle & 0 & 0 \\
      \lambda_{u,31}\langle H \rangle & 0 & 0 
    \end{array}
  \right)
  \label{final1}
\end{equation}
Using independent non-singular transformations on the $Q_{i}$ and
$u_{i}$ fields, one can find bases in which eq.~\eqref{final1} becomes
\begin{equation}
  \left(
    \begin{array}{ccc}
      0 &    0       &    0\\
      0 & \lambda\langle H \rangle &    0 \\
      0 &    0       & \lambda\langle H \rangle
    \end{array}
  \right)
  \label{final2}
\end{equation}
where $\lambda$ is an arbitrary, but non-zero, real number. We
conclude from the zero diagonal element that one up-quark is strictly
massless\footnote{At least, on the classical level. Higher order and
  non-perturbative terms in the superpotential could lead to naturally
  small corrections.}.  Furthermore, the two non-zero diagonal
elements imply that the second and third up-quarks will have
non-vanishing masses of $O\big(\langle H \rangle\big)$. However, the
exact value of their masses will depend on the explicit normalization
of the kinetic energy terms in the low energy theory. These masses,
therefore, are in general not degenerate.  This analysis applies to
the down-quarks and the up- and down-leptons as well.  We conclude
that, prior to higher order and non-perturbative corrections, one
complete generation of quarks/leptons will be massless.  The remaining
two generations will have non-vanishing masses on the order of the
electroweak symmetry breaking scale which are, generically,
non-degenerate.

The coefficients $\lambda$ have no interpretation as an intersection
number and, therefore, no reason to be constant over the moduli space.
In general, we expect them to depend on the moduli. Of course, to
explicitly compute the quark/lepton masses one needs, in addition, the
K\"ahler potential, which determines the correct normalization of the
fields.

\section*{Acknowledgments}

We are grateful to R.~Donagi, B.~Nelson, and T.~Pantev for
enlightening discussions. This research was supported in part by the
Department of Physics and the Math/Physics Research Group at the
University of Pennsylvania under cooperative research agreement
DE-FG02-95ER40893 with the U.~S.~Department of Energy and an NSF
Focused Research Grant DMS0139799 for ``The Geometry of
Superstrings.'' T.~P.~is partially supported by an NSF grant DMS
0403884. Y.-H.~H.~is also indebted to the FitzJames Fellowship of
Merton College, Oxford.

\bibliographystyle{JHEP} \renewcommand{\refname}{Bibliography}
\addcontentsline{toc}{section}{Bibliography} \bibliography{Yukawa}

\providecommand{\href}[2]{#2}\begingroup\raggedright\begin{thebibliography}{10}

\bibitem{Binetruy:2005ez}
P.~Binetruy, G.~L. Kane, J.~D. Lykken, and B.~D. Nelson, {\it Twenty-five
  questions for string theorists},
  \href{http://xxx.lanl.gov/abs/hep-th/0509157}{{\tt hep-th/0509157}}.

\bibitem{HetSM1}
V.~Braun, Y.-H. He, B.~A. Ovrut, and T.~Pantev, {\it A heterotic standard
  model},  \href{http://xxx.lanl.gov/abs/hep-th/0501070}{{\tt hep-th/0501070}}.

\bibitem{HetSM2}
V.~Braun, Y.-H. He, B.~A. Ovrut, and T.~Pantev, {\it A standard model from the
  {$E_8 \times E_8$} heterotic superstring},
  \href{http://xxx.lanl.gov/abs/hep-th/0502155}{{\tt hep-th/0502155}}.

\bibitem{HetSM3}
V.~Braun, Y.-H. He, B.~A. Ovrut, and T.~Pantev, {\it Vector bundle extensions,
  sheaf cohomology, and the heterotic standard model},
  \href{http://xxx.lanl.gov/abs/hep-th/0505041}{{\tt hep-th/0505041}}.

\bibitem{MinimalHetSM}
V.~Braun, Y.-H. He, B.~A. Ovrut, and T.~Pantev, {\it The exact {MSSM} spectrum
  from string theory},  \href{http://xxx.lanl.gov/abs/hep-th/0512177}{{\tt
  hep-th/0512177}}.

\bibitem{Gomez:2005ii}
T.~L. Gomez, S.~Lukic, and I.~Sols, {\it Constraining the {K\"ahler} moduli in
  the heterotic standard model},
  \href{http://xxx.lanl.gov/abs/hep-th/0512205}{{\tt hep-th/0512205}}.

\bibitem{future}
``Stability.'' to appear.

\bibitem{Gross:1985rr}
D.~J. Gross, J.~A. Harvey, E.~J. Martinec, and R.~Rohm, {\it Heterotic string
  theory. 2. the interacting heterotic string},  {\em Nucl. Phys.} {\bf B267}
  (1986) 75.

\bibitem{Sen:1985eb}
A.~Sen, {\it The heterotic string in arbitrary background field},  {\em Phys.
  Rev.} {\bf D32} (1985) 2102.

\bibitem{Evans:1985vb}
M.~Evans and B.~A. Ovrut, {\it Splitting the superstring vacuum degeneracy},
  {\em Phys. Lett.} {\bf B174} (1986) 63.

\bibitem{Breit:1985ud}
J.~D. Breit, B.~A. Ovrut, and G.~C. Segre, {\it {$E_6$} symmetry breaking in
  the superstring theory},  {\em Phys. Lett.} {\bf B158} (1985) 33.

\bibitem{Aspinwall:1987cn}
P.~S. Aspinwall, B.~R. Greene, K.~H. Kirklin, and P.~J. Miron, {\it Searching
  for three generation {C}alabi-{Y}au manifolds},  {\em Nucl. Phys.} {\bf B294}
  (1987) 193.

\bibitem{Green:1987mn}
M.~B. Green, J.~H. Schwarz, and E.~Witten, {\it Superstring theory. vol. 2:
  Loop amplitudes, anomalies and phenomenology}, . Cambridge, Uk: Univ. Pr. (
  1987) 596 P. ( Cambridge Monographs On Mathematical Physics).

\bibitem{Curio:1998vu}
G.~Curio, {\it Chiral matter and transitions in heterotic string models},  {\em
  Phys. Lett.} {\bf B435} (1998) 39--48,
  [\href{http://xxx.lanl.gov/abs/hep-th/9803224}{{\tt hep-th/9803224}}].

\bibitem{Andreas:1999ty}
B.~Andreas, G.~Curio, and A.~Klemm, {\it Towards the standard model spectrum
  from elliptic {C}alabi-{Y}au},  {\em Int. J. Mod. Phys.} {\bf A19} (2004)
  1987, [\href{http://xxx.lanl.gov/abs/hep-th/9903052}{{\tt hep-th/9903052}}].

\bibitem{Donagi:1999gc}
R.~Donagi, A.~Lukas, B.~A. Ovrut, and D.~Waldram, {\it Holomorphic vector
  bundles and non-perturbative vacua in {M}- theory},  {\em JHEP} {\bf 06}
  (1999) 034, [\href{http://xxx.lanl.gov/abs/hep-th/9901009}{{\tt
  hep-th/9901009}}].

\bibitem{Krause:2000gp}
A.~Krause, {\it A small cosmological constant, grand unification and warped
  geometry},  \href{http://xxx.lanl.gov/abs/hep-th/0006226}{{\tt
  hep-th/0006226}}.

\bibitem{Andreas:2003zb}
B.~Andreas and D.~Hernandez-Ruiperez, {\it Comments on {$N=1$} heterotic string
  vacua},  {\em Adv. Theor. Math. Phys.} {\bf 7} (2004) 751--786,
  [\href{http://xxx.lanl.gov/abs/hep-th/0305123}{{\tt hep-th/0305123}}].

\bibitem{Curio:2003ur}
G.~Curio and A.~Krause, {\it Enlarging the parameter space of heterotic
  {M}-theory flux compactifications to phenomenological viability},  {\em Nucl.
  Phys.} {\bf B693} (2004) 195--222,
  [\href{http://xxx.lanl.gov/abs/hep-th/0308202}{{\tt hep-th/0308202}}].

\bibitem{Curio:2004pf}
G.~Curio, {\it Standard model bundles of the heterotic string},
  \href{http://xxx.lanl.gov/abs/hep-th/0412182}{{\tt hep-th/0412182}}.

\bibitem{Blumenhagen:2005ga}
R.~Blumenhagen, G.~Honecker, and T.~Weigand, {\it Loop-corrected
  compactifications of the heterotic string with line bundles},
  \href{http://xxx.lanl.gov/abs/hep-th/0504232}{{\tt hep-th/0504232}}.

\bibitem{Blumenhagen:2005zg}
R.~Blumenhagen, G.~Honecker, and T.~Weigand, {\it Non-abelian brane worlds: The
  heterotic string story},  \href{http://xxx.lanl.gov/abs/hep-th/0510049}{{\tt
  hep-th/0510049}}.

\bibitem{Becker:2005sg}
K.~Becker, M.~Becker, and A.~Krause, {\it {M}-theory inflation from multi
  {M5}-brane dynamics},  {\em Nucl. Phys.} {\bf B715} (2005) 349--371,
  [\href{http://xxx.lanl.gov/abs/hep-th/0501130}{{\tt hep-th/0501130}}].

\bibitem{Raby:2005vc}
S.~Raby, {\it Three family models from the heterotic string},
  \href{http://xxx.lanl.gov/abs/hep-th/0510014}{{\tt hep-th/0510014}}.

\bibitem{SM-bundle1}
R.~Donagi, B.~A. Ovrut, T.~Pantev, and D.~Waldram, {\it Standard-model bundles
  on non-simply connected {Calabi-Yau} threefolds},  {\em JHEP} {\bf 08} (2001)
  053, [\href{http://xxx.lanl.gov/abs/hep-th/0008008}{{\tt hep-th/0008008}}].

\bibitem{SM-bundle2}
R.~Donagi, B.~A. Ovrut, T.~Pantev, and D.~Waldram, {\it Standard-model
  bundles},  {\em Adv.Theor.Math.Phys.} {\bf 5} (2002) 563--615,
  [\href{http://xxx.lanl.gov/abs/math.AG/0008010}{{\tt math.AG/0008010}}].

\bibitem{SM-bundle3}
R.~Donagi, B.~A. Ovrut, T.~Pantev, and D.~Waldram, {\it Spectral involutions on
  rational elliptic surfaces},  {\em Adv.Theor.Math.Phys.} {\bf 5} (2002)
  499--561, [\href{http://xxx.lanl.gov/abs/math.AG/0008011}{{\tt
  math.AG/0008011}}].

\bibitem{z2z2-1}
B.~A. Ovrut, T.~Pantev, and R.~Reinbacher, {\it Invariant homology on {Standard
  Model} manifolds},  {\em JHEP} {\bf 01} (2004) 059,
  [\href{http://xxx.lanl.gov/abs/hep-th/0303020}{{\tt hep-th/0303020}}].

\bibitem{z2z2-2}
R.~Donagi, B.~A. Ovrut, T.~Pantev, and R.~Reinbacher, {\it ${SU(4)}$ instantons
  on {Calabi-Yau} threefolds with {$\Z_2 \times \Z_2$} fundamental group},
  {\em JHEP} {\bf 01} (2004) 022,
  [\href{http://xxx.lanl.gov/abs/hep-th/0307273}{{\tt hep-th/0307273}}].

\bibitem{dP9Z3Z3}
V.~Braun, B.~A. Ovrut, T.~Pantev, and R.~Reinbacher, {\it Elliptic
  {{Calabi-Yau}} threefolds with {$\Z_3 \times \Z_3$} {W}ilson lines},  {\em
  JHEP} {\bf 12} (2004) 062,
  [\href{http://xxx.lanl.gov/abs/hep-th/0410055}{{\tt hep-th/0410055}}].

\bibitem{DonagiPrincipal}
R.~Y. Donagi, {\it Principal bundles on elliptic fibrations},  {\em Asian J.
  Math.} {\bf 1} (1997), no.~2 214--223,
  [\href{http://xxx.lanl.gov/abs/alg-geom/9702002}{{\tt alg-geom/9702002}}].

\bibitem{FMW1}
R.~Friedman, J.~Morgan, and E.~Witten, {\it Vector bundles and {F} theory},
  {\em Commun. Math. Phys.} {\bf 187} (1997) 679--743,
  [\href{http://xxx.lanl.gov/abs/hep-th/9701162}{{\tt hep-th/9701162}}].

\bibitem{FMW2}
R.~Friedman, J.~Morgan, and E.~Witten, {\it Principal {G}-bundles over elliptic
  curves},  {\em Math.Res.Lett.} {\bf 5} (1998) 97--118,
  [\href{http://xxx.lanl.gov/abs/alg-geom/9707004}{{\tt alg-geom/9707004}}].

\bibitem{FMW3}
R.~Friedman, J.~Morgan, and E.~Witten, {\it Vector bundles over elliptic
  fibrations},  \href{http://xxx.lanl.gov/abs/alg-geom/9709029}{{\tt
  alg-geom/9709029}}.

\bibitem{Diaconescu:1998kg}
D.-E. Diaconescu and G.~Ionesei, {\it Spectral covers, charged matter and
  bundle cohomology},  {\em JHEP} {\bf 12} (1998) 001,
  [\href{http://xxx.lanl.gov/abs/hep-th/9811129}{{\tt hep-th/9811129}}].

\bibitem{Donagi:2004qk}
R.~Donagi, Y.-H. He, B.~A. Ovrut, and R.~Reinbacher, {\it Moduli dependent
  spectra of heterotic compactifications},
  \href{http://xxx.lanl.gov/abs/hep-th/0403291}{{\tt hep-th/0403291}}.

\bibitem{Donagi:2004ia}
R.~Donagi, Y.-H. He, B.~A. Ovrut, and R.~Reinbacher, {\it The particle spectrum
  of heterotic compactifications},
  \href{http://xxx.lanl.gov/abs/hep-th/0405014}{{\tt hep-th/0405014}}.

\bibitem{Ovrut:2002jk}
B.~A. Ovrut, T.~Pantev, and R.~Reinbacher, {\it Torus-fibered {Calabi-Yau}
  threefolds with non-trivial fundamental group},  {\em JHEP} {\bf 05} (2003)
  040, [\href{http://xxx.lanl.gov/abs/hep-th/0212221}{{\tt hep-th/0212221}}].

\bibitem{MR1797016}
R.~P. Thomas, {\it An obstructed bundle on a {C}alabi-{Y}au 3-fold},  {\em Adv.
  Theor. Math. Phys.} {\bf 3} (1999), no.~3 567--576.

\bibitem{MR1807601}
R.~P. Thomas, {\it Examples of bundles on {C}alabi-{Y}au 3-folds for string
  theory compactifications},  {\em Adv. Theor. Math. Phys.} {\bf 4} (2000),
  no.~1 231--247.

\bibitem{SU5-z2-1}
R.~Donagi, Y.-H. He, B.~A. Ovrut, and R.~Reinbacher, {\it Higgs doublets, split
  multiplets and heterotic {$SU(3)_C \times SU(2)_L \times U(1)_Y$} spectra},
  {\em Phys. Lett.} {\bf B618} (2005) 259--264,
  [\href{http://xxx.lanl.gov/abs/hep-th/0409291}{{\tt hep-th/0409291}}].

\bibitem{SU5-z2-2}
R.~Donagi, Y.-H. He, B.~A. Ovrut, and R.~Reinbacher, {\it The spectra of
  heterotic standard model vacua},  {\em JHEP} {\bf 06} (2005) 070,
  [\href{http://xxx.lanl.gov/abs/hep-th/0411156}{{\tt hep-th/0411156}}].

\bibitem{ModHetSM}
V.~Braun, Y.-H. He, B.~A. Ovrut, and T.~Pantev, {\it Heterotic standard model
  moduli},  \href{http://xxx.lanl.gov/abs/hep-th/0509051}{{\tt
  hep-th/0509051}}.

\bibitem{Bouchard:2005ag}
V.~Bouchard and R.~Donagi, {\it An {$SU(5)$} heterotic standard model},
  \href{http://xxx.lanl.gov/abs/hep-th/0512149}{{\tt hep-th/0512149}}.

\bibitem{Buchmuller:2005jr}
W.~Buchmuller, K.~Hamaguchi, O.~Lebedev, and M.~Ratz, {\it The supersymmetric
  standard model from the heterotic string},
  \href{http://xxx.lanl.gov/abs/hep-ph/0511035}{{\tt hep-ph/0511035}}.

\bibitem{HiggsMuTerm}
V.~Braun, Y.-H. He, B.~A. Ovrut, and T.~Pantev, {\it Moduli dependent
  {$\mu$}-terms in a heterotic standard model},
  \href{http://xxx.lanl.gov/abs/hep-th/0510142}{{\tt hep-th/0510142}}.

\bibitem{Giedt:2005vx}
J.~Giedt, G.~L. Kane, P.~Langacker, and B.~D. Nelson, {\it Massive neutrinos
  and (heterotic) string theory},
  \href{http://xxx.lanl.gov/abs/hep-th/0502032}{{\tt hep-th/0502032}}.

\bibitem{Greene:1987xh}
B.~R. Greene, K.~H. Kirklin, P.~J. Miron, and G.~G. Ross, {\it {$27^3$}
  {Yukawa} couplings for a three generation superstring model},  {\em Phys.
  Lett.} {\bf B192} (1987) 111.

\bibitem{Gepner:1988ck}
D.~Gepner, {\it Yukawa couplings for {Calabi-Yau} string compactification},
  {\em Nucl. Phys.} {\bf B311} (1988) 191.

\end{thebibliography}\endgroup

\end{document}